# The Distant Heart: Mediating Long-Distance Relationships through Connected Computational Jewelry

Yulia Silina, Hamed Haddadi
Queen Mary University of London

## ABSTRACT
In the world where increasingly mobility and long-distance relationships (LDR) with family, friends and loved-ones became commonplace, there exists a gap in intimate interpersonal communication mediated by technology. Considering the advances in the field of mediation of relationships through technology, as well prevalence of use of jewelry as love-tokens for expressing a wish to be remembered and to evoke the presence of the loved-one, developments in the new field of *computational jewelry* offer some truly exciting possibilities.

In this paper we investigate the role that the jewelry-like form factor of prototypes can play in the context of studying effects of *computational jewelry* in mediating LRD. We perform a usability survey, then design, build, and evaluate a wearable display of one's heart rate, shared remotely using the Twitter network. We do this through developing jewelry-like prototype using a multidisciplinary design-driven approach within *Human-Centered Design* [3] framework and newly established strategies for mediating intimate relationships trough technology [1].

## Author Keywords
Computational jewelry; jewelry-like prototypes; mediating relationships through technology; Human-centered design; wearable technologies; intimacy; presence.

## ACM Classification Keywords
H.5.m. Information interfaces and presentation (e.g., HCI): Miscellaneous.

## INTRODUCTION
The LDR are a prevalent cause of stress and can generate a profound sense of longing and nostalgia. It is common to use a variety of methods to stay in touch with each other. But despite wide availability of audio, video and written communications, it is not unusual for us to experience occasions when all we want is to share a moment with the loved-ones and let them know that we are thinking of them without having to say or write any words. It is equally important to know that the other person cares and thinks about one, even if they do not necessarily directly communicate such emotions in voice or writing.

One of the cultural practices that allow individuals to intimately and meaningfully conjure up memories and presence of each other is an ancient practice of exchanging love-tokens and keepsakes such as wedding rings and pendent lockets. As pervasive as this practice is, the actual objects are static and afford little interaction. Once the object is gifted, the giver has no influence on its state and the wearer of the object has to rely on the imagination and memories to conjure up the presence of the giver. Nevertheless, this practice endures [5] despite the growth of modern communication technologies.

The rapid development of the new multidisciplinary field of *computational jewelry* [2] (colloquially known as *smart* or *digital jewelry*) is on a brink of merging technological potential of Smart Watches and self-quantifying gadgets with aesthetic and cultural significance of jewelry. This practice would enable creating connected love-tokens and keepsakes that could facilitate implicit communication and expression within the existing cultural framework.

In this paper we investigate the appropriateness of the jewelry-like form factor in prototypes, used in research about effects *connected computational jewelry* has in mediating the emotional gap created by LDR.

## RELATED WORK
There is distinct rise of *computational jewelry* in fashion [10] and a growing body of research in the field of mediation of intimate relationships through technology [1, 8], but the exact role of connected *computational jewelry* as a mediator of relationships is yet to be fully understood.

Unfortunately, most research into *computational jewelry* is done using working prototypes at best. This puts evaluating users in an unnatural scenario where they attempt intimate communication whist being "decorated" with gadgets, circuit boards and wires. Even the researches that coined the term *computation jewelry* [2] overlook the importance of collaboration with jewelers in creating prototypes with compelling jewelry-like form factor.

| | |
|---|---|
| **62%** | participants were **women** |
| **92%** | between the ages of 21-50 (**47** % of **31-40**) |
| **72%** | **employed** |
| **77%** | love or **like technology** |
| **72%** | **would share heartbeat** if they could |
| **100%** | **own a smart phone** |
| **100%** | **live away** from someone they loved |

Figure 1. Key points about the questionnaire participants

We propose that in order to minimize disruption when studying effects of *computational jewelry* as a mediator of relationships, it is important to make prototypes in a jewelry-like form-factor that would suite the users [8]. Thus, we aim to create and test in situ a device that felt and looked like decorative jewelry, familiar to its users, whilst fulfilling its mediation and computational functions.

## DEVELOPMENT

### Design Approach

Taking into account dual nature of *computational jewelry* as both jewelry and a computational device, the research team drew on the multidisciplinary expertise. On one side, the practical experience of jewelry design and making and in depth knowledge of current fashion allowed us to insure that the prototype was in line with the current trends and user-expectation of the jewelry. On the other, side, we applied an overarching *Human-Centered Design* (HCD) framework [3], widely used in HCI research and practice as well as in creative practices and jewelry design.

In addition to the HCD framework, we applied a recently developed toolkit for *Current Strategies of Mediating Intimate Relationships through Technology* [2]. To determine a specific HCD solution we applied *Iterative Design Cycle* [3], evaluating each phase in the context of *"Do…? Feel…? Know..?"* framework [7]. This was done through evaluating a set of questionnaires, scenarios, cultural references, prototypes, heuristics, as well as practicalities of designing and making jewelry.

### Conceptual Model

We conceived *The Distant Heart* system, in homage to the seminal *connected computational necklace*, designed by Moi Sompit Fusakul for the *Cyborg II* Experiment [9]. As such, *The Distant Heart* system was intended to include two people that love and miss each other and that are located at a distance for a prolonged period of time during the day. One person (*Sharer*) would measure and share biometric data. The data would be received by the necklace and revealed to the other person (*Necklace Wearer*), enhancing their sense of connectedness [4].

Our particular study was not intended to focus on the effects of the specific biometric input. Thus to represent Fusakul's original idea of biological emotional indicator, we fell back on a recent work [6] indicating that the heartbeat is a potent symbol of intimate connection of people in LDR. Heartbeat was also relevant as a familiar cultural icon, widely used in keepsakes and love-tokens.

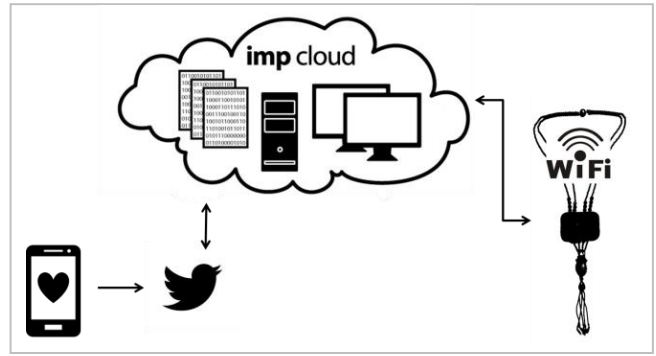

Figure 2. Infrastructure of The Distant Heart

### Establishing the Users

To establish potential users for *The Distant Heart* system, along with their goals, tastes and objectives, we devised a web-based semi-structured questionnaire that asked general demographic questions, along with the questions about relationships with loved-ones that live away, the characteristics of loved-ones that are missed, levels of privacy of manifestations of these characteristics and so on.

The questionnaire was distributed though social media with snowball sampling and offered no financial rewards. We received 100 responses at different levels of completion. Out of these, 66 were fully complete and informed our choices in the development of *The Distant Heart*.

The results (Figure 1) indicated that the *Necklace Wearer* falls within the commercial precedent of *computational jewelry* that is aimed at women between 21-50 years old. Further to that, the *Wearer* would likely be professional, interested in and familiar with technology and wishing to receive the heartbeat of the romantic partner when he or she is away. Out of six mediation strategies [2], the survey results indicated that *physicalness* (heartbeat), *awareness* and *expressivity* were most important.

### System Infrastructure

The system was setup where the *Sharer* could use any off-the shelf heart-rate measuring mobile app[1] that had a connection to a dedicated *Twitter* account. The necklace was operated and connected to WiFi using an *Electric Imp*[2], which has efficient battery consumption, allowing at least 24-hour of usage.

The necklace's *Agent*, located at the *Imp Cloud* periodically scanned a dedicated (protected) Twitter account for new data. If shared heartbeat was detected, the *Imp Cloud* sent a signal back to the necklace (Figure 2). The necklace then notified the *Necklace Wearer* through a short sequence of vibrations that their loved-one shared a heartbeat with them. The necklace then expressed the heartbeat, unless the *Wearer* chose not to see the expression by pressing the OFF button.

---

[1] Most users chose *What's My Heart Rate* by ViTrox or similar

[2] www.electricimp.com

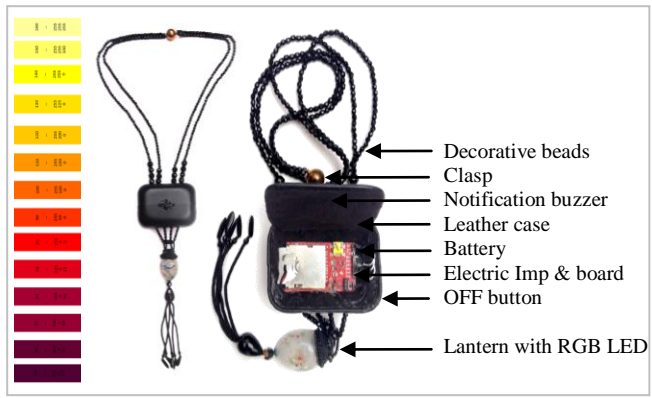

**Figure 3. The Distant Heart necklace components**

### Necklace Design

The heartbeat was represented by the ambient color-changing, with frequency of the heartbeat corresponding to the frequency of pulsation of the light. The RGB values were mapped to colors on the warm spectrum of the color wheel with red-purple representing heart rate at rest, and bright yellow heart rate after exercises with the gradation of colors associated with the heart rate in-between (Figure 3).

To prevent the "techno-bling" effect and to stay in keeping with the cultural references of love-tokens and current fashion, a set of traditional materials and techniques were employed in making of the necklace. The final necklace looked like a ready-to wear necklace, consisting of the double-string of black Swarovski crystal beads, a fashionable 3cm x 4.5cm leather case in a form of oriental central bead and a miniature egg-shaped frosted glass lantern with the painted images, commonly associated with tenderness and love through many cultures (Figure 3).

## EVALUATION

### Methodology

To evaluate if the prototype, had a desired effect on bridging the emotional gap between the two people that miss each other, we conducted a set of user studies. Each user study lasted a day, during which we asked pairs of participants to be away from each other for at least 6 hours.

The *Necklace Wearers* were met in-situ, explained how the necklace worked, and left to their regular chores. The *Sharers* were located elsewhere and were instructed via e-mail. Informal observations were made before and after each study to insure that the user study went smoothly. During each study, we recorded and logged the usage data. Upon completion, both *Sharers* and *Wearers* were asked to fill in an individual questionnaire, relating their experience of using the system.

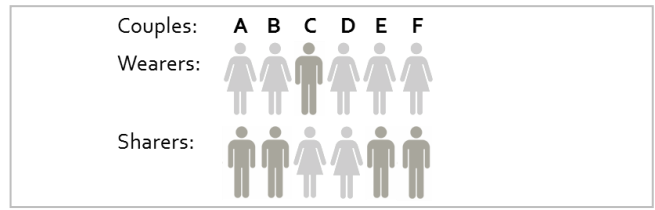

**Figure 4. Participants of the user study**

### User Study Participants

Users were invited to participate in the study through the initial web-survey and through additional social media channels with snowball sampling and no financial reward. Out of 37 people who expressed the interested to participate in the study 14 people (7 couples) were logistically able to participate. One couple failed to complete the questionnaire, despite giving positive verbal feedback. Thus, findings below include 6 couples, in line with studies in similar areas of research.

As anticipated by the results of initial questionnaire, participation in the user study of *The Distant Heart* was instigated by women, who wanted to see a heartbeat of their loved-one, or by the men who were more interested in the technology. Participants were between 21 and 50, engaged with technology, with all but one *Wearers* being female, and most *Sharers* male (Figure 4).

All couples were involved in romantic relationship between three weeks and ten years. Couple F reported to be engaged in a prolonged long-distance relationship. Couple D was apart for several days, and the rest were apart for the duration of the user study. Couple A chose to use *The Distant Heart* necklace on their wedding anniversary. Couples A, B, E, and F used *The Distant Heart* system during the work week, and C and D on the weekend.

### Results and Analysis

Overall all participants enjoyed the use of *The Distant Heart* and found the idea interesting. When asked what they thought initially about it, two distinct sentiments emerged which echoed results of questionnaire: *"Well, it's quite romantic. Cyberpunk romantic"* (Sharer B), and *"We've had a long distance relationship before; we're also both interested in innovative uses of technology."* (Sharer A).

When relating the experience in the questionnaire, female participants describe emotions, whereas male described technology. This is likely because mainly male *Sharers* used off-the-shelf technology during the user study. Anticipating this, we expected that *Wearers* might find the experience unalike other forms of communication, were as *Sharers* would find the experience the same as usual. To our surprise, 7 out of 12 participants indicated that they had a special experience that would not be replicated by other means of communications. And 8 out of 12 participants indicated that using *The Distant Heart* system made them feel closer to their loved-one and that they would use this technology if it was available (Figure 5). In response to the questions about how soon they anticipate the "novelty factor" to wear out, the reply was between few days and a few months. To remedy, participants suggested using this system on special occasions when their loved-one is away.

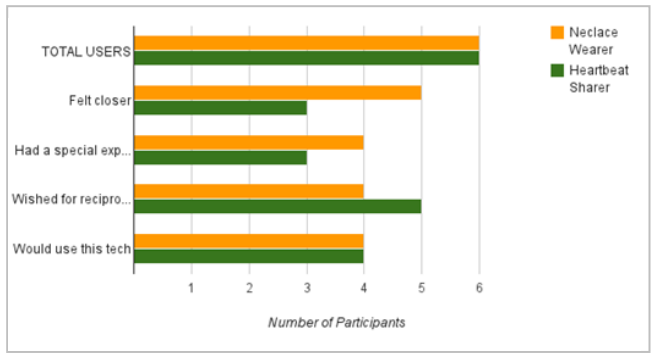

**Figure 5. Sentiment of The Distant Heart use**

During the use of *The Distant Heart*, each couple shared and received between 5 and 13 heartbeats. Interestingly, Sharer E sent one of the most numbers of heartbeats and was sentimental about the use of the system; whereas Wearer E had a strong opinion that the system is unnecessary and she is perfectly happy to convey her emotions though regular forms of mobile communication.

None of the *Wearers* chose to terminate broadcast of the heartbeat, indicating late that it did not disturb their chores. Some participants pointed out a few inconveniences: such as sensitivity and power-consumption of heart rate measuring app and slight heaviness of the necklace. However, these did not seem to affect overall sentiment negatively. Participants particularly liked that the interaction was intimate and did not involve any words. Among others, Wearer B expressed that: *"it was free of words. a very strange/awesome combination of technology and under-the-skinness",* and Sharer D, expressed is as *"direct intimacy, no symbols or signs or language."*

All participants had a strong preference for reciprocity, which the system did not allow. *Necklace Wearers* in particular reported that they would have liked to be able to request the heartbeat from their loved-one. Wearer E made a poignant comment: *"It was lovely too see and feel that my boyfriend is thinking about me, and actually takes the effort to measure his heartbeat and send it to me. It was strange though, that i could not reply or react to this anyhow.."*

When asked what they thought about the device being in a form of jewelry, participants responded positively. For example, Wearer B said that *"It makes it more magical. A very good approach. Would definitely not want (or have) the same experience with something super sleek looking. It would feel like a sporting equipment then. Aka would not have the same emotional intensity."*

## OPPORTUNITIES & FUTURE WORK

These initial results are encouraging and suggest that designing the prototypes in jewelry-form factor is useful. In the future, it might be interesting to develop and test a more sophisticated system that addresses reciprocity. There are also indications of the potential use of connected *computational jewelry* with various interaction modalities for bringing together not only romantic couples, but also friends, family members, and even owners of pets. The web-based questionnaire, the user study and informal observations indicate that women are most likely wearers of connected *computational jewelry*. But there might also be a further split between the gender and age groups in their interests and preferences related to *computation jewelry*.

## CONCLUSION

*The Distant Heart* necklace was created through multidisciplinary approach, using *HCD* framework [3] and newly emerged mediation strategies [2]. The conducted user study demonstrated that most *Necklace Wearers* and at least a half of *Sharers* had a special intimate experience and felt more connected when using *The Distant Heart* necklace. These early findings, coupled with the enduring historical use of the jewelry in a form of love-tokens, support the importance of multidisciplinary approach in creating prototypes that felt and looked like decorative jewelry, familiar to its users, whilst fulfilling its mediation and computational functions and that could be tested in situ.